# Constructing the Best Trading Strategy:
# A New General Framework

*Philip Z. Maymin and Zakhar G. Maymin*

**Abstract**: We introduce a new general framework for constructing the best trading strategy for a given historical indicator. We construct the unique trading strategy with the highest expected return. This optimal strategy may be implemented directly, or its expected return may be used as a benchmark to evaluate how far away from the optimal other proposed strategies for the given indicators are. Separately, we also construct the unique trading strategy with the highest information ratio. In the normal case, when the traded security return is near zero, and for reasonable correlations, the performance differences are economically insignificant. However, when the correlation approaches one, the trading strategy with the highest expected return approaches its maximum information ratio of 1.32 while the trading strategy with the highest information ratio goes to infinity.

Philip Z. Maymin (corresponding author)
NYU-Polytechnic Institute
Six MetroTech Center
Brooklyn, NY 11201
Telephone: 718-260-3175
Email: phil@maymin.com

Zakhar G. Maymin
Quantitative Investment Services, Inc.
Email: zak@maymin.com

# Introduction

There is a vast literature analyzing the performance of strategies formed from various historical indicators comparing different techniques, implementations, and sets of parameters, including momentum and mean-reversion strategies (cf. Jegadeesh and Titman 2002, Conrad and Kaul 1998), fundamentals (cf. Fama and French 2008, Pesaran and Timmermann 1995), technical analysis (cf. Brock, Lakonishok, and LeBaron 1992, Faber 2007), and machine learning (cf. Enke and Thawornwong 2005, Kim 2003). However, it has always been difficult to determine with any certainty whether a particular technique or a particular strategy could have performed better with different parameters or different implementations. Further, for the strategies that have been most successful, it is unclear how much they could have been further improved, if at all.

This paper aims to answer these questions. We develop a new general framework for analyzing strategies based on a given indicator, use that framework to find the greatest expected return with the best possible strategy, construct such a strategy explicitly, and derive the implications.

The best strategy could be interpreted as the one that has the highest expected return or the highest information ratio. (Considering the information instead of the Sharpe ratio does not conceptually make any difference but simplifies the presentation.) In this paper we will discuss and compare both.

The framework assumes that we know the conditional distribution of the future returns of the traded securities, given the historical indicator. As with the fields of mean-variance efficient portfolio allocation and option and derivatives pricing theory, such an assumption has its advantages and disadvantages. Among the advantages are the ability to quickly compare the potential profitability of trading for various indicators, markets, and securities and the ability to estimate how far from the best a particular trading strategy is. Thus, this approach helps better allocate resources in developing a trading strategy. The disadvantages of such an approach are that the true conditional distribution is unknown and often not stable in financial markets. This paper does not address time-varying estimation issues of the conditional distribution, though we do discuss simple ways to estimate the optimal strategy from the data.

Even if the conditional distribution is completely known, a trader may still prefer to use another, non-optimal strategy if, for example, the execution of the sub-optimal strategy is easier, or if it incurs less slippage, commissions, or other transactions costs. In such cases, our framework quantifies how far from optimal that strategy lies and gives the trader a quick tool to do performance analysis.

Any developer of a trading strategy could benefit from our new framework in three ways. First, our framework allows a more robust analysis and comparison of strategies across different indicators and different markets. Second, our framework provides a quick metric by which the distance to the optimal strategy can be measured. Third, our framework overcomes the problem of overfitting data because it requires only a handful of intuitively understandable parameters such as volatilities and correlation.

## Framework

Let us define a *trading strategy* as a pair of a *historical indicator H*, and a *notional function $f = f(H)$*, a function on $H$. Let's also define $R$ as the next period return of the traded security.

The historical indicator $H$ could be a function of the historical prices of the traded security or a function of any historical information on any securities, markets, indices, factors, or anything else. The role of the indicator is to aggregate historical information before investing in the security. An example of an indicator in the case of momentum might be the 12-month trailing return on the security.

The notional function $f = f(H)$ tells us how much we should invest in the security after observing the indicator $H$, or how many shares or contracts or other units of the security we should buy if $f > 0$, or sell if $f < 0$. To reflect capital constraints, we will assume that $f$ is a bounded function.

Often, there are different amounts of money required to make a trade depending on whether we buy or sell the security. For example, typically there are different margin requirements for long and short positions in stock or long and short positions in options, or constraints or restrictions on borrowing that preclude selling short. Also, because of risk considerations, sometimes the margin depends on the composition of the current portfolio to which we add this security. Yet all the results of this paper are trivially generalized for the case when $f(H)$ is bounded above and below by different constants; thus, without loss of generality, we'll assume that the magnitude of $f(H)$ is bounded by 1, so that $|f| \leq 1$. This means that we are investing a maximum of $1 for the next time period. Let us call such strategies *notional-constrained*.

Typically $R$ is the return we get during the next time period, or excess return, but it could also be the profit and loss (i.e., the dollar amount we receive or pay), the volatility, or any other variable for which a market exists that we could buy or sell. It could be a discrete or even binary variable, such as binary bets. In addition, the time period of the future results need not be a fixed time period but could also be defined relative to other occurrences (e.g., stop losses). Here, without loss of generality, we consider $R$ to be the next period return of the security.

Thus the return of our strategy on our investment during the next time period is:

$$Q \stackrel{\text{def}}{=} f(H) \cdot R$$

Now we can formulate our main questions: investing not more than $1, what is the maximum expected return (or information ratio) of the optimal strategy and what notional function $f(H)$ maximizes it?

We first find and analyze the trading strategy with the highest expected return for a single indicator and a single security. We then generalize the results for multiple indicators and multiple securities and explore the implications. Finally, we compare the results with the trading strategy having the highest information ratio.

# The Trading Strategy with the Highest Expected Return

## Theorem

Let the historical indicator $H$ and the return $R$ be random variables with a known conditional expectation $g(H) \stackrel{\text{def}}{=} E(R|H)$. And let's consider only notional functions $f(H)$ such that $|f(H)| \leq 1$.

Then the maximum expected return

$$\max_{|f| \leq 1} E(f(H)R) = E(|g(H)|)$$

is achieved only by the optimal notional function

$$f^*(H) = \text{sign}(g(H))$$

The optimal notional function is unique up to the set $H_0 = \{h: g(h) = 0\}$.

## Proof

The idea of the proof is to show that we cannot improve the optimal signal $f^*(h)$ on any infinitesimal interval.

Let $Q = f(H)R$ as above be the strategy return of an arbitrary notional $f(H)$, $|f(H)| \leq 1$, and let $\delta_h(H) = 1$ for $h \leq H \leq h + dh$, and 0 elsewhere, for any real $h$ and $dh > 0$.

Then, for an $\varepsilon > 0$, and $dh \to 0$,

$$E\big[(f(H) + \varepsilon\,\delta_h(H))R\big] = E[f(H)g(H) + \varepsilon\,g(H)\delta_h(H)]$$

$$= E\big[f(H)g(H)(1 - \delta_h(H))\big] + E[f(H)g(H)\delta_h(H)] + E[\varepsilon\,g(H)\delta_h(H)]$$

$$\approx E\big[f(H)g(H)(1 - \delta_h(H))\big] + (f(h) + \varepsilon)g(h)dh$$

with the first equality following from the fact that the unconditional expected value of a random variable is the unconditional expected value of the conditional expected value of that variable given any indicator.

Therefore, when $g(h) > 0$, to be optimal, $f(h)$ must be exactly 1, or else we would be able to improve it at the point $h$ by adding a small enough $\varepsilon$ to keep $f(h) \leq 1$. Similarly, when $g(h) < 0$, $f(h)$ must be $-1$. When $g(h) = 0$, the value of $f(h)$ doesn't affect the strategy expected return.

## The Normal Case

Let the historical indicator $H$ and return $R$ be bivariate normal random variables with $EH = \mu_H, Var(H) = \sigma_H^2, ER = \mu, Var(R) = \sigma^2$, and $Corr(H, R) = \rho$. Without loss of generality, we assume that $\rho > 0$. For negative correlations, we can redefine the indicator to have the opposite sign. The special case of zero correlation is treated in the next subsection.

Denote by $A(\theta)$ the expectation of the absolute value of a normal distribution with mean $\theta$ and variance 1:

$$A(\theta) \stackrel{\text{def}}{=} E(|X|), X \sim \mathcal{N}(\theta, 1) = e^{-\frac{\theta^2}{2}}\sqrt{2/\pi} + \theta(2\mathcal{N}(\theta) - 1)$$

Then the maximum expected return of the optimal strategy is:

$$M(\mu, \sigma, \rho) \stackrel{\text{def}}{=} \max_{|f| \leq 1} E[f(H)R] = E(|E(R|H)|) = \rho\sigma A\left(\frac{\mu}{\rho\sigma}\right) = \rho\sigma A(m)$$

where $\mathcal{N}(\cdot)$ is the CDF of the standard normal distribution and $m$ is the "m-ratio" to be discussed below:

$$m \stackrel{\text{def}}{=} \frac{\mu}{\rho\sigma}$$

Note that the maximum expected return $M$ does not depend on the distribution parameters $\mu_H$ and $\sigma_H$ of the indicator.

The maximum is achieved only by the optimal notional function:

$$f^*(H) = \text{sign}(E(R|H)) = \text{sign}\left(\rho\sigma\left(\frac{H - \mu_H}{\sigma_H}\right) + \mu\right) = \text{sign}(\rho\sigma H^* + \mu)$$

Note that the optimal notional function will be unchanged if the historical indicator is standardized to $H^* = (H - \mu_H)/\sigma_H$. Thus, without loss of generality, we can assume that $\mu_H = 0$ and $\sigma_H = 1$.

In the normal case, the optimal notional is equal to the sign of the regression of the return $R$ on the indicator $H$. This follows because the conditional expected value for normal variables is the regression value of one variable on the others.

With the simplifying assumptions that the historical indicator has been standardized, and that its correlation with the return is positive, we can rewrite the optimal notional function as follows:

$$f^*(H) = \text{sign}(H + m)$$

There are two cases. If $\mu \geq 0$, then if the historical indicator is positive or zero, the optimal trading strategy buys, and continues buying until the indicator falls below a given threshold. If $\mu \leq 0$, then if the historical indicator is negative or zero, the optimal trading strategy sells, and continues selling until the indicator exceeds a given threshold. In each case the threshold is $-m$, the opposite of the m-ratio.

What is the m-ratio? It is the security's information ratio divided by its correlation with the indicator, and it can also be calculated as the ratio of the expected return to the expected product of the indicator and the return:

$$m = \frac{\mu}{\rho\sigma} = \frac{E(R)}{E(HR)}$$

In practice, this may serve as a useful way to estimate $-m$, the threshold of optimal strategies.

**No Knowledge**

In the normal case, if the historical indicator provides no knowledge about the future return, then the correlation between them is zero, and the conditional expected value of $R$ given $H$ is the same as the unconditional expected value of $R$. Therefore, the optimal strategy in the case of zero knowledge is to buy and hold the security if its expected return is positive, and to sell it otherwise.

**Perfect Knowledge**

In the normal case, if the historical indicator provides perfect knowledge about the future return, then the correlation between them is one, and the best strategy is the one that buys when it is known that the return will be positive and sells when it is known that the return will be negative. Hence the expected return of the optimal strategy will simply be the expected value of the absolute value of $R$:

$$A\left(\frac{\mu}{\sigma}\right)\sigma$$

For the typical case in practice of $R$ representing excess returns with expected value $\mu = 0$, the optimal expected return is $A(0)\sigma = \sigma\sqrt{2/\pi} \approx 0.8\sigma$.

### Negligible Drift

As the security's expected return approaches zero, the expected return of the optimal trading strategy approaches:

$$\rho \sigma A(0) = \rho \sigma \sqrt{2/\pi} \approx 0.8 \rho \sigma$$

Thus in the case of negligible drift for the security, the best possible expected return is simply proportional in the correlation to the perfect knowledge case, and the optimal notional function is:

$$f^*(H) = \text{sign}(H)$$

In this case, the return of the optimal strategy is more likely to be positive than negative. In fact, we can calculate an exact expression for the probability that the return of the optimal strategy exceeds zero. Let $Q^* = f^*(H)R$ be the return of the optimal strategy. Then:

$$\Pr(Q^* \geq 0) = \frac{1}{2} + \frac{\text{ArcSin}(\rho)}{\pi}$$

Figure 1 plots this probability as a function of the correlation $0 \leq \rho \leq 1$. The probability equals one-half when $\rho$ is zero, always exceeds one-half for positive $\rho$, and is equal to one when $\rho$ is one.

Thus, any strategy on a security with negligible drift that is more likely to be negative than positive is not the optimal strategy.

**Figure 1: Probability of Positive Return of Optimal Strategy**
This figure plots the probability that the return of the optimal strategy is positive as a function of the correlation between the security and the historical indicator.

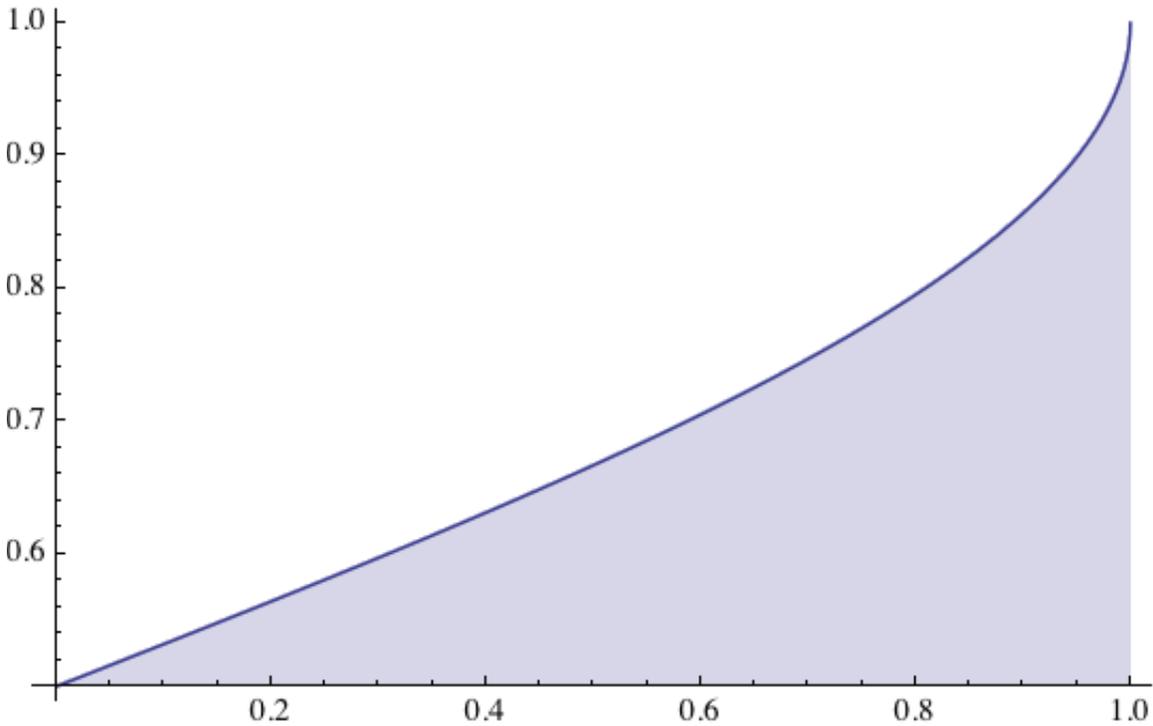

**Relation with the m-Ratio**

In the general case, the optimal expected return is $\rho\sigma A(m)$. If we fix $\rho\sigma$ and let $\mu = m\rho\sigma$ vary, then we can compare different strategies on an equal footing, and for a single security, we can then evaluate the sensitivity of the strategy to changes in the security's drift.

Thus, we can plot $A(m)$, the optimal expected return as a proportion of $\rho\sigma$. Figure 2 graphs $A(m)$ and $|m|$ versus $m$. When $m = 0$, we obtain the performance of the negligible drift, $A(0) \approx 0.8$. As $m$ moves away from zero, the performance improves. Once the absolute value of $m$ exceeds one, the difference between $A(m)$ and $|m|$ becomes negligible: $A(1) = 1.167, A(2) = 2.017, A(3) = 3.001$.

**Figure 2:** $A(m)$ **and** $|m|$

This figure plots $A(m)$, the optimal expected return as a proportion of $\rho\sigma$, and $|m|$, the absolute value of $m$, versus $m$.

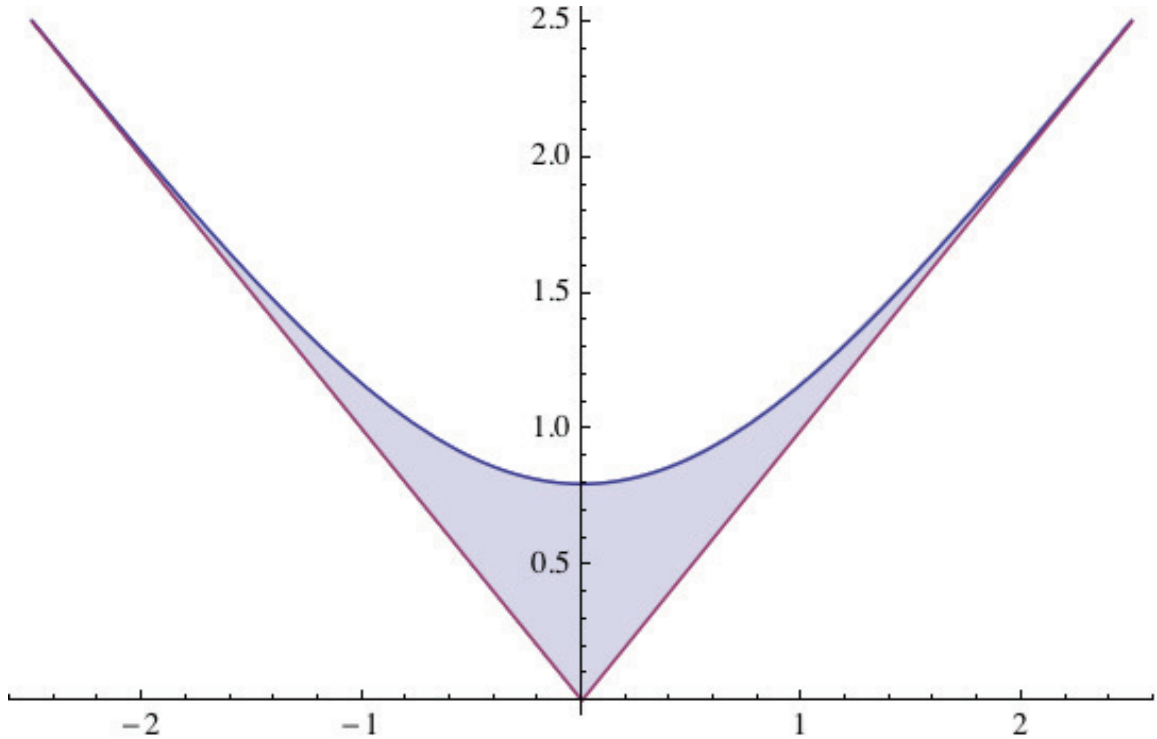

## Examples and Extensions

### Non-Linear Conditional Expected Value of $R$ on $H$

The main result above says that the best notional function is given by the sign of the conditional expected value of the return $R$ on the historical indicator $H$. In the case of the bivariate normal, and in many other bivariate distributions, the conditional expected value is linear in $R$. In practice, however, when we plot $R$ against $H$, we can sometimes see a non-linear relation.

To estimate conditional expected values we could use non-linear regressions of returns on indicators. Copulas are another possible approach.

Copulas are useful for such situations, because they provides examples of non-linear families of conditional expected values based only on a few parameters, often based only on one parameter. Crane and van der Hoek (2008) provide multiple examples of such copulas.

Based on the plot of $R$ against $H$, the appropriate family of copulas can be chosen for which the parameters can be estimated from the data to evaluate the conditional expected value of $R$ given $H$. Then the main result can be used to estimate the best notional.

### Multiple Indicators

If instead of a single historical indicator $H$ we have multiple indicators $H_1, H_2, \ldots, H_n$, then analogously the best notional function is:

$$f^*(H_1, H_2, \ldots, H_3) = \text{sign}\big(E(R|H_1, H_2, \ldots, H_3)\big)$$

The proof is similar to that of the main result.

Having multiple indicators is effectively equivalent to having a single historical indicator $H_0 = E(R|H_1, H_2, \ldots, H_3)$. For normally distributed indicators and returns, the single indicator will simply be a linear combination of the other indicators, the result of the relevant regression.

### Multiple Securities

#### Theorem

Let historical indicators $\boldsymbol{H} = H_1, H_2, \ldots, H_n$ and returns $R_1, R_2, \ldots, R_k$ be random variables with a known joint distribution. And let's consider only notional functions $f(\boldsymbol{H}) = f(H_1, H_2, \ldots, H_n)$ that are bounded by 1.

Then the maximum expected return is achieved only by the following optimal notional function:

$$f^*(\mathbf{H}) = \text{sign}\left(E(R_{j^*}|H)\right)$$

where $j^*$ is such that:

$$E(\text{sign}(E(R_{j^*}|H))R_{j^*}) = \max_{j=1,2,\ldots,k} E(\text{sign}(E(R_j|H))R_j)$$

In other words, the optimal trading strategy would in each instance invest only in the security with the highest absolute value of expected returns.

In practice, however, it may be beneficial to invest in several securities, especially if the expected returns of applying the optimal strategies to them are similar, to lower the portfolio volatility.

## Optimal Strategy Information Ratio

### Derivation of General Formula

As above, denote by $Q$ the return of the optimal strategy:

$$Q \stackrel{\text{def}}{=} f^*(H)R$$

The main result for the bivariate normal distribution shows that:

$$f^*(H) = \text{sign}(g)$$

where $g \stackrel{\text{def}}{=} E(R|H) = \rho\sigma H + \mu$ and $M = E(Q)$ as above.

We can compute the variance of $Q$ as:

$$V \stackrel{\text{def}}{=} Var(Q) = E(Q^2) - M^2 = E(R^2) - M^2 = \sigma^2 + \mu^2 - M^2$$

with the last equality following because $\text{sign}(g)^2 = 1$.

Thus the information ratio of the optimal strategy is:

$$\Omega \stackrel{\text{def}}{=} \frac{M}{\sqrt{V}} = \frac{e^{-m^2/2}\left(\sqrt{2} + \sqrt{\pi}\tau m e^{m^2/2}\right)}{\sqrt{\pi/\rho^2 + \pi m^2 + \pi\tau^2 m^2 - 2\sqrt{2\pi}\tau m e^{-m^2/2} - 2e^{-m^2}}}$$

where $m = \frac{\mu}{\rho\sigma}$ is the m-ratio as above and $\tau = \tau(m) \stackrel{\text{def}}{=} 2\mathcal{N}(m) - 1$.

Note that the information ratio depends only on the m-ratio and the correlation; equivalently, it is depends only on the correlation and the security's information ratio $\omega \stackrel{\text{def}}{=} \mu/\sigma$.

Thus we can view the strategy information ratio as:

$$\Omega = \Omega(\omega, \rho)$$

The information ratio of the optimal strategy will always equal or exceed the information ratio of the security, even when the distribution is not normal, because, as we have shown earlier, the expected return of the optimal strategy will always equal or exceed the expected return of any particular strategy, for example the buy-and-hold strategy, and therefore it will always exceed the expected return of the security:

$$E(Q^*) = E(f^*(H)R) \geq E(R) > 0$$

and because the second moment of the security will always equal or exceed the second moment of the optimal strategy:

$$E(R^2) \geq E\left(f^{*2}(H)R^2\right) = E(Q^{*2})$$

with strict inequality holding if there exists at least one value of $H = h$ for which $f^*(h) = 0$, and so the variance of the optimal strategy will never exceed the variance of the security.

**Large Security Information Ratio**

Figure 3 plots the information ratio of the optimal strategy for various values of the information ratio of the security and various levels of correlation between the security and the historical indicator.

The optimal strategy information ratio approaches the security information ratio as the latter goes to infinity:

$$\lim_{\omega \to \infty} \Omega/\omega = 1$$

Asymptotically, the plot converges quickly to $|\omega| = |\mu/\sigma|$, as can be seen from the table at the bottom of Figure 3.

Essentially, this means that the best strategy for securities with large information ratios cannot substantially outperform the buy-and-hold strategy.

**Figure 3: Optimal Strategy Information Ratio**
This figure shows the information ratio of the optimal strategy $\Omega$ as a function of the security information ratio $\omega$ and the correlation $\rho$ between the security return and the historical indicator. The table beneath it selects a few points to illustrate the speed of convergence.

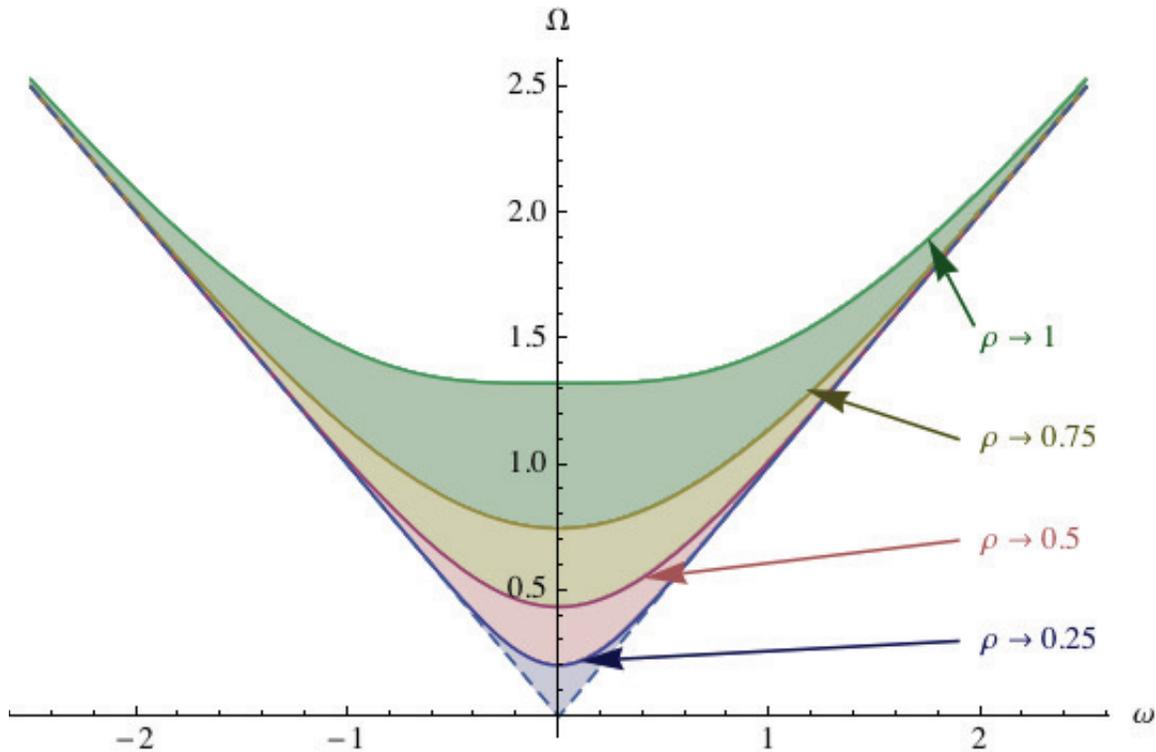

| $\rho$ | $\Omega(\omega =.5, \rho)$ | $\Omega(\omega = 1, \rho)$ | $\Omega(\omega = 2, \rho)$ |
|---|---|---|---|
| 0.1 | 0.5 | 1. | 2. |
| 0.25 | 0.505324 | 1.00001 | 2. |
| 0.5 | 0.611567 | 1.0172 | 2.00004 |
| 0.75 | 0.85525 | 1.1411 | 2.00891 |
| 1. | 1.33818 | 1.45946 | 2.08951 |

### Negligible Security Information Ratio

When the security information ratio $\omega$ is near zero, the optimal strategy information ratio $\Omega$ simplifies to:

$$\Omega(0,\rho) = \frac{\sqrt{2}\rho}{\sqrt{\pi - 2\rho^2}}$$

Figure 4 plots this function, which increases from zero when the correlation is zero to a maximum of 1.324 when the correlation is one.

Another way of deriving this maximum is to note that it is the information ratio of a strategy with perfect knowledge trading one unit of the security. Such a strategy would earn the absolute value of the security return, $|R|$, with mean

$$A(0)\sigma = \sqrt{2/\pi}\sigma \approx 0.8\sigma$$

and variance

$$\sigma^2(1 - A(0)^2)$$

so its information ratio would be

$$\frac{A(0)}{\sqrt{1 - A(0)^2}} \approx 1.324$$

Annualizing, the information ratio of the strategy with the highest expected return from monthly trading on an indicator with perfect knowledge cannot exceed $1.324\sqrt{12} \approx 4.6$. Strategies purporting to exceed this limit must either have suboptimal expected return, or trade more than one unit of the security, or trade more than one security.

**Figure 4: Optimal Strategy Information Ratio for a Zero-Drift Security**
This figure plots, for a security with zero drift, the information ratio of the optimal strategy $\Omega(0, \rho)$ as a function of the correlation $\rho$ between the security return and the historical indicator.

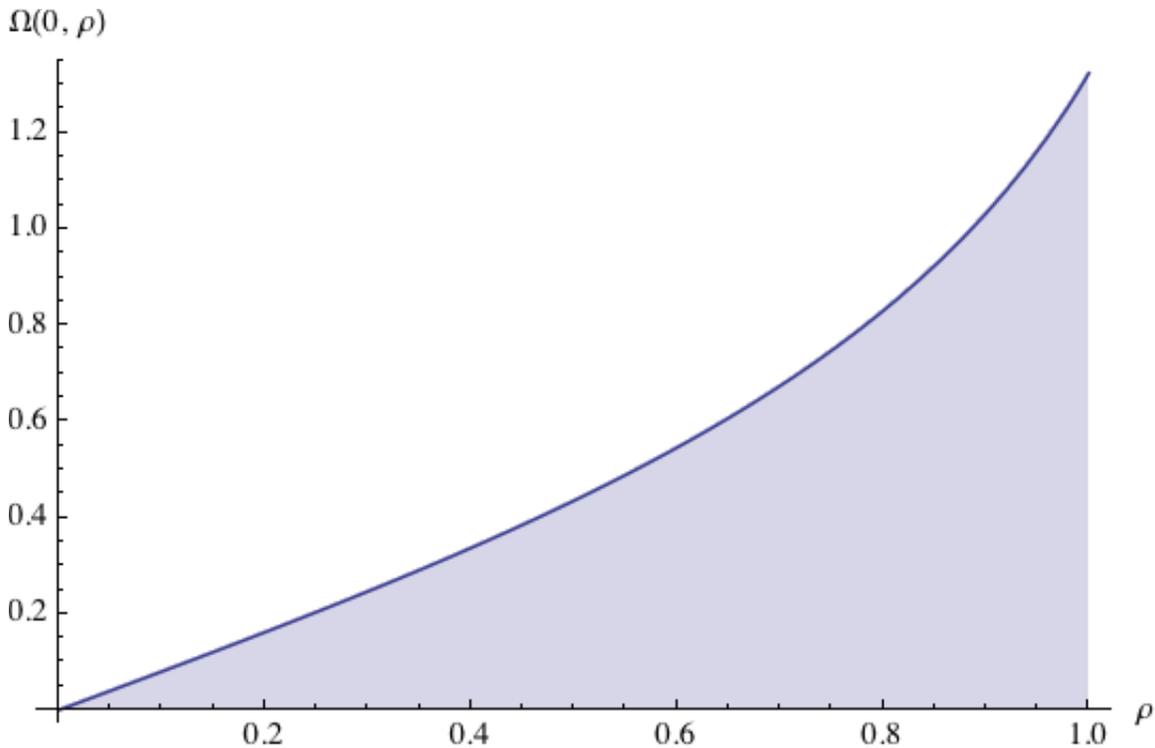

**Relation with Market Efficiency**

Following cf. Fama (1976), market efficiency cannot be tested without a model of market equilibrium. One common approach to testing market efficiency is to compare the returns on a particular strategy with the returns from buying and holding the security itself. If the strategy returns statistically significantly exceed the security returns, inefficiency is claimed.

The implications from our approach are that particular strategies no longer need to be developed and tested for the purpose of testing market efficiency. Once a historical indicator correlated with future returns is found, then we have shown there will exist an optimal strategy with expected returns in excess of the security returns. Moreover, the information ratio of the optimal strategy will exceed the information ratio of the security.

Another way of seeing the relation with market efficiency is as follows. We have shown in full generality, assuming as usual that the security has positive expected return $E(R) > 0$, that:

$$E(R) \leq E\big(\text{sign}\big(E(R|H)\big)R\big) \leq E(|R|)$$

The first inequality is an equality if and only if $E(R|H) = E(R)$. In the language of Fama (1976), this means the market is efficient. In other words, in the context of testing the market for inefficiency conditional on a given indicator, rather than attempting to continually develop and improve strategies to see if their expected return can outperform buy-and-hold, our approach is able to directly construct the optimal strategy. This removes the possibility that the failure to prove inefficiency was caused by a poor implementation of a strategy.

## The Trading Strategy with the Highest Information Ratio

### Theorem

Let the indicator $H$ and the return $R$ again be random variables with known first two conditional moments $g_1(H) = E(R|H)$ and $g_2(H) = E(R^2|H)$, and let:

$$g(H) \stackrel{\text{def}}{=} g_1(H)/g_2(H)$$

$$g(H,\lambda) \stackrel{\text{def}}{=} \psi\big(\lambda g_1(H)/g_2(H)\big)$$

$$\zeta = E(g_1^2/g_2) = E(gR)$$

where $\psi(\cdot)$ is the clipping function:

$$\psi(x) \overset{\text{def}}{=} \begin{cases} -1, & x \leq -1 \\ x, & |x| < 1 \\ 1, & x \geq 1 \end{cases}$$

Finally, as before, a trading strategy is notional-constrained if $|f(H)| \leq 1$.

Then:

1. The maximum information ratio across all possible strategies is:

$$\frac{\sqrt{\zeta}}{\sqrt{1-\zeta}}$$

   and this maximum is achieved with the notional function $f^*(H) = g(H)$.

2. If $g(H)$ is bounded or $H$ is bounded, $|g(H)| \leq C$ for all $H$, then the maximum information ratio across all notional-constrained strategies is

$$\frac{\sqrt{\zeta}}{\sqrt{1-\zeta}}$$

   and this maximum is achieved with the notional function $f^*(H) = g(H)/C$.

3. The maximum information ratio across all notional-constrained strategies is:

$$\max_{\lambda > 0} \frac{E(g(H,\lambda)g_1(H))}{\sqrt{E(g(H,\lambda)g_2(H))^2 - \left(E\big(g(H,\lambda)g_1(H)\big)\right)^2}}$$

   and the notional function $g(H, \lambda)$ achieves the greatest return among all notional-constrained strategies with the same second moment as $g(H, \lambda)$.

**Proof**

1. The information ratio of a strategy's return $f(H)R$ for a notional function $f(H)$ is

$$\frac{E(fR)}{\sqrt{E(f^2R^2) - (E(fR))^2}} = \frac{E(fg_1)}{\sqrt{E(f^2g_2) - (E(fg_1))^2}}$$

Thus $f(H)$ maximizes the information ratio if and only if it maximizes:

$$\frac{(E(fg_1))^2}{E(f^2g_2)}$$

Suppose a strategy with notional function $f(H)$ has a greater information ratio than that of $f^*(H)$ and denote by $C$ the second moment of its strategy return:

$$E(f^2R^2) = E(f^2g_2) = C, C > 0$$

Define

$$\lambda \stackrel{\text{def}}{=} \frac{\sqrt{E(g^2g_2)}}{2\sqrt{C}} > 0$$

Then for any $H$, $f_\lambda \stackrel{\text{def}}{=} \frac{g_1}{2\lambda g_2} = \frac{g}{2\lambda}$ is the particular $f$ that maximizes

$$fg_1 - \lambda(f^2g_2)$$

and

$$E(f_\lambda^2 g_2) = E\left(\left(\frac{g}{2\lambda}\right)^2 g_2\right) = \frac{1}{4\lambda^2} E(g^2 g_2) = \frac{(2\lambda\sqrt{C})^2}{4\lambda^2} = C$$

So for any $f = f(H)$ and all $H$:

$$f_\lambda g_1 - \lambda(f_\lambda^2 g_2) \geq fg_1 - \lambda(f^2g_2)$$

and

$$E(f_\lambda g_1) - \lambda E(f_\lambda^2 g_2) \geq E(fg_1) - \lambda E(f^2 g_2)$$

So

$$E(f_\lambda g_1) \geq E(f g_1)$$

Therefore the information ratio of $f_\lambda$ is greater than that of $f$. But the information ratio of $f_\lambda$ equals the information ratio of $f^* = g = 2\lambda f_\lambda$ because multiplying the notional by a positive constant does not change a strategy's information ratio.

Thus $f^* = g$ is the optimal trading strategy having the highest information ratio, and its information ratio is:

$$\frac{E(gg_1)}{\sqrt{E(g^2 g_2) - (E(gg_1))^2}} = \frac{E(g_1^2/g_2)}{\sqrt{E(g_1^2/g_2) - (E(g_1^2/g_2))^2}} = \frac{\sqrt{\zeta}}{\sqrt{1-\zeta}}$$

Furthermore, $\zeta \leq 1$ because $(E(R|H))^2 \leq E(R^2|H)$ and with equality holding if and only in the case of perfect knowledge of $H$ on $R$.

2. The proof of this statement follows immediately from the preceding statement because again multiplying a notional function by a constant does not affect the strategy's information ratio and we know that $f^* = g$ has the greatest information ratio among all possible strategies, including notional-constrained strategies. Further, if $H$ is bounded and $g(H)$ is continuous, then $g(H)$ must also be bounded, and in practice, $H$ is always bounded.

3. The proof of this statement (3) is similar to the proof of statement (1). This statement is useful in situations when $g$ is unbounded to find the notional-constrained strategy with the best information ratio by looking at notional-constrained strategies having a fixed second moment.

**Normal Case**

As before, let the historical indicator $H$ and return $R$ be bivariate normal random variables with $EH = \mu_H$, $Var(H) = \sigma_H^2$, $ER = \mu$, $Var(R) = \sigma^2$, and $Corr(H, R) = \rho > 0$.

Then:

$$g_1(H) = \mu + \frac{H - \mu_H}{\sigma_H} \rho \sigma$$

$$g_2(H) = \frac{\mu^2 \sigma_H^2 + 2\rho(H - \mu_H)\mu\sigma_H\sigma + (\rho^2(H - \mu_H)^2 + (1 - \rho^2)\sigma_H^2)\sigma^2}{\sigma_H^2}$$

$$g(H) = \frac{\sigma_H(\mu\sigma_H + \rho(H - \mu_H)\sigma)}{\mu^2\sigma_H^2 + 2\rho(H - \mu_H)\mu\sigma_H\sigma + (\rho^2(H - \mu_H)^2 + (1 - \rho^2)\sigma_H^2)\sigma^2}$$

As above, only the standardized version of $H$ enters these equations. Therefore, without loss of generality, we can assume in what follows that $\mu_H = 0$ and $\sigma_H = 1$, or, equivalently, that the historical indicator $H$ is standardized.

Because $\rho$ is positive, $g(H)$ is obviously bounded and therefore the optimal information ratio notional function is

$$f^*(H) = g(H)/g(H_+)$$

where $H_+$ and $H_-$ are such that

$$g(H_+) = \max_H g(H) = \frac{1}{2\sqrt{1 - \rho^2}\sigma}$$

$$g(H_-) = \min_H g(H) = -\frac{1}{2\sqrt{1-\rho^2}\sigma}$$

and

$$H_\pm = -m \pm \sqrt{\frac{1-\rho^2}{\rho^2}}$$

where as above

$$m \stackrel{\text{def}}{=} \frac{\mu}{\rho\sigma}$$

So

$$f^*(H) = \frac{2\sqrt{1-\rho^2}\sigma(\mu + H\rho\sigma)}{\mu^2 + (1-\rho^2)\sigma^2 + 2H\mu\rho\sigma + H^2\rho^2\sigma^2}$$

and

$$\zeta = \int_{-\infty}^{\infty} \frac{e^{-\frac{H^2}{2}}(\mu + H\rho\sigma)^2}{\sqrt{2\pi}(\mu^2 + (1-\rho^2)\sigma^2 + 2H\mu\rho\sigma + H^2\rho^2\sigma^2)} dH$$

so the maximum information ratio is as above:

$$\frac{\sqrt{\zeta}}{\sqrt{1-\zeta}}$$

and the expected return of the maximum information ratio strategy is:

$$E(f^*R) = \frac{E(gR)}{g(H_+)} = 2\sqrt{1-\rho^2}\sigma\zeta$$

## Normal Case with Negligible Drift

Suppose now that $\mu = 0$. Then:

$$f^*(H) = \frac{2\rho\sqrt{1-\rho^2}H}{1-\rho^2+\rho^2 H^2}$$

$$\zeta = 1 - \sqrt{2\pi}e^{\frac{1}{2}b^2}b(1-\mathcal{N}(b))$$

where

$$b = b(\rho) \stackrel{\text{def}}{=} \frac{\sqrt{1-\rho^2}}{\rho}$$

and the maximum information ratio is:

$$\frac{\sqrt{\zeta}}{\sqrt{1-\zeta}}$$

Figure 5 plots the optimal notional function $f^*$ for a variety of correlations. It is a symmetric function around zero, is equal to zero for $H = 0$, and is bounded by +1 and -1. As the correlation approaches one, the maximum occurs closer to $H = 0$ and quickly approaches zero elsewhere. As the correlation approaches zero, the notional function approaches $f^*(H) = 0$.

Figure 6 plots the expected return of the optimal information ratio strategy. For a correlation of zero, the expected return approaches zero. For a correlation of one, the expected return is 100%.

## Figure 5: Optimal Notional Function for Negligible Drift

This figure plots the notional function with the highest information ratio in the normal case for a negligible drift for a variety of correlations, as well as sign($H$) which is the notional function with the highest expected return.

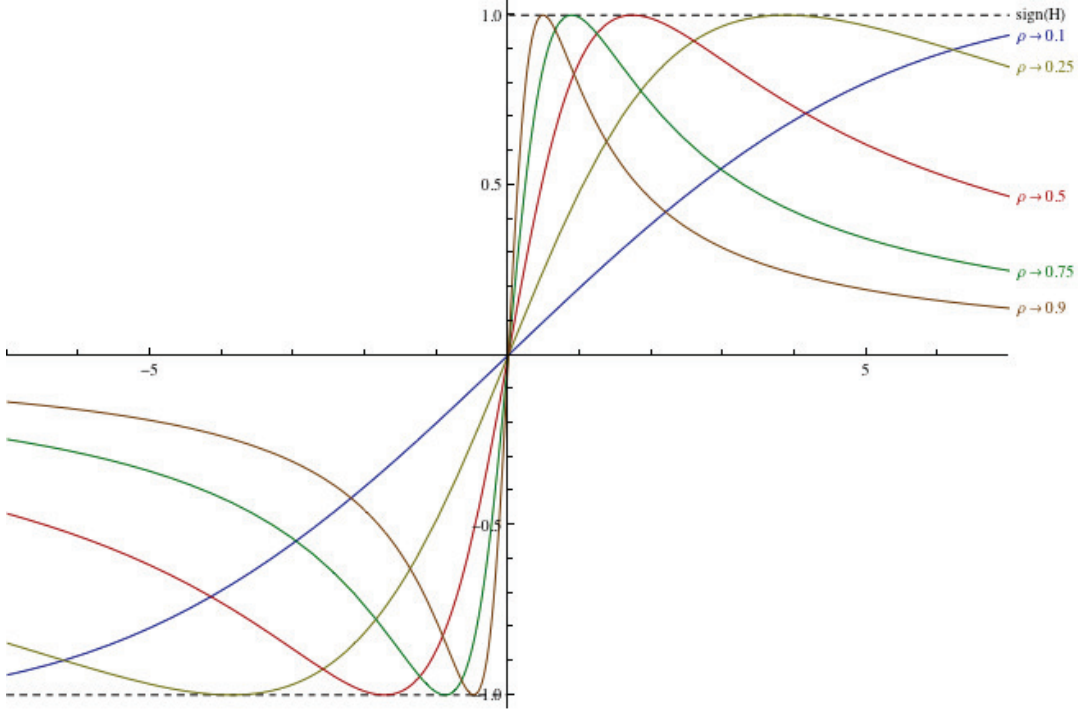

## Figure 6: Expected Return of Optimal Information Ratio Strategy

This figure plots the expected return $\zeta$ of the strategy with the highest information ratio as a function of the correlation $\rho > 0$.

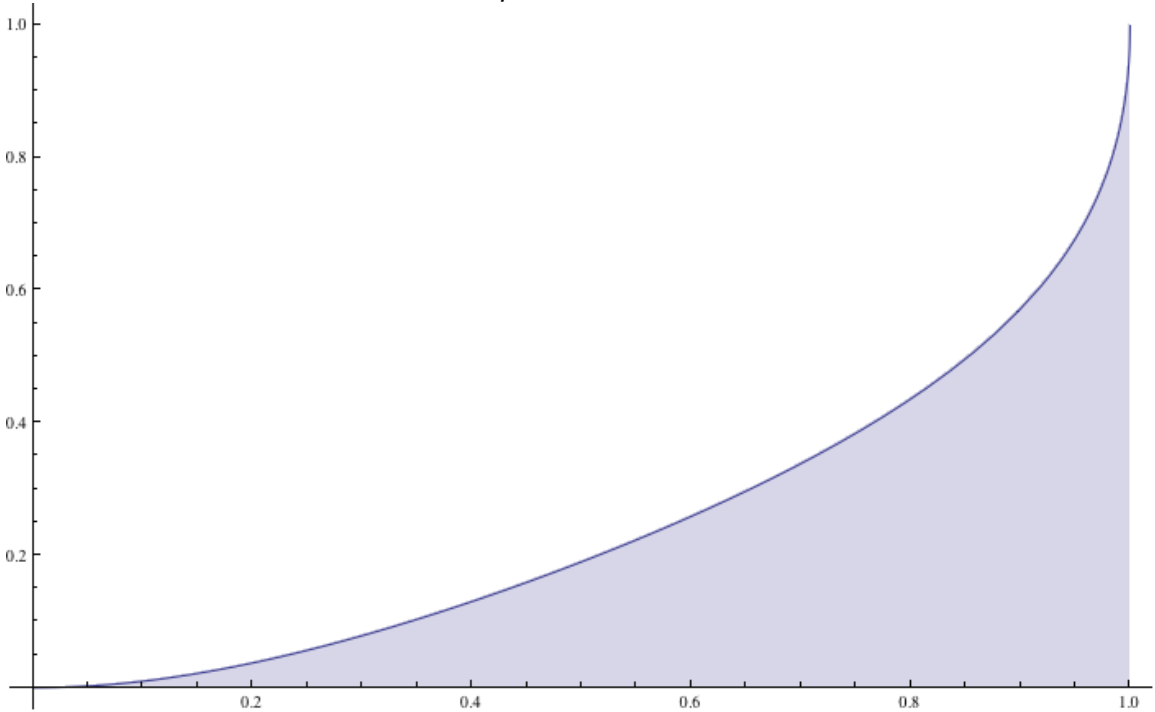

### Figure 7: Maximum Information Ratio

This figure plots the maximum information ratio of the optimal strategy as a function of the correlation $\rho$.

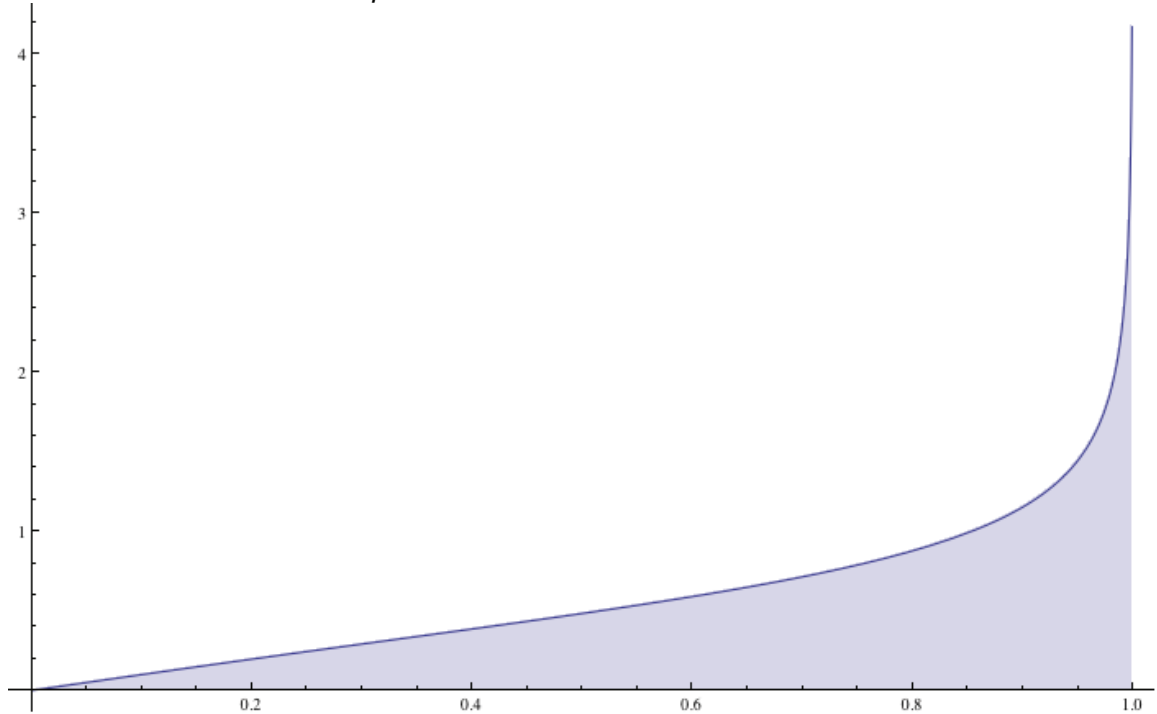

| Correlation $\rho$ | Maximum Information Ratio $\sqrt{\zeta}/\sqrt{1-\zeta}$ | Approximation for $\rho \to 1$ $(\pi(1-\rho))^{-1/4}$ |
|---|---|---|
| 0.1 | 0.0995317 | 0.771173 |
| 0.2 | 0.196865 | 0.794219 |
| 0.3 | 0.291882 | 0.821179 |
| 0.4 | 0.386508 | 0.853443 |
| 0.5 | 0.484213 | 0.893244 |
| 0.6 | 0.590425 | 0.94449 |
| 0.7 | 0.714714 | 1.01492 |
| 0.8 | 0.878467 | 1.12319 |
| 0.9 | 1.15429 | 1.33571 |
| 0.99 | 2.29156 | 2.37527 |
| 0.999 | 4.17963 | 4.22389 |
| 0.9999 | 7.48685 | 7.51126 |
| 0.99999 | 13.3435 | 13.3571 |

Figure 7 plots the information ratio of the optimal strategy, equal to $\sqrt{\zeta}/\sqrt{1-\zeta}$. The maximum information ratio grows as $\rho$ until $\rho$ exceeds about 0.7, and then as $\rho$ goes to one, the maximum information ratio goes to infinity as:

$$\left(\pi(1-\rho)\right)^{-1/4}$$

The table at the bottom of Figure 7 displays the maximum information ratio for several correlation values. It also displays the approximation formula above.

### Comparison with Highest Expected Return Strategy

Let's compare the strategy with the highest information ratio to the strategy with the highest expected return. Table 1 summarizes the notional function, expected return, standard deviation, and information ratio for each of the two types.

### Table 1: Summary of Optimal Strategies
This table summarizes the optimal notional function, the expected return of the strategy, its standard deviation, and its information ratio, for the maximum expected return strategy and the maximum information ratio strategy.

|  | Maximum Expected Return Strategy | Maximum Information Ratio Strategy |
|---|---|---|
| **Optimal Notional** $f^*$ | $\text{sign}(H)$ | $\dfrac{2\rho\sqrt{1-\rho^2}H}{1-\rho^2+\rho^2 H^2}$ |
| **Expected Return** $E(f^*R)$ | $\rho\sigma\sqrt{2/\pi}$ | $2b\rho\sigma\zeta$ |
| **Standard Deviation** $SD(f^*R)$ | $\sqrt{1-2\rho^2/\pi}\,\sigma$ | $2b\rho\sigma\sqrt{\zeta}\sqrt{1-\zeta}$ |
| **Information Ratio** $E(f^*R)/SD(f^*R)$ | $\dfrac{\sqrt{2}\rho}{\sqrt{\pi-2\rho^2}}$ | $\sqrt{\zeta}/\sqrt{1-\zeta}$ |

where
$$\zeta = 1 - \sqrt{2\pi}e^{\frac{1}{2}b^2}b(1-\mathcal{N}(b))$$
and
$$b \stackrel{\text{def}}{=} \sqrt{1-\rho^2}/\rho$$

Each optimal strategy's expected return is proportional to the security volatility $\sigma$. Figure 8 plots the ratio of the expected return of the optimal expected return strategy to that of the optimal information ratio strategy. This ratio goes to infinity as the correlation goes either to zero or to one, and otherwise it always exceeds one, achieving a minimum of 1.15 when $\rho = 0.65$. This ratio can be expressed as:

$$\frac{\sqrt{e}\rho^2}{\sqrt{2e\pi}\rho\sqrt{1-\rho^2} - 2e^{\frac{1}{2\rho^2}}\pi(1-\rho^2)\left(1 - \mathcal{N}(\sqrt{1-\rho^2}/\rho)\right)}$$

**Figure 8: The Ratio of Expected Returns for the Optimal Strategies**
This figure plots the ratio of expected returns for the optimal expected return strategy to the optimal information ratio strategy.

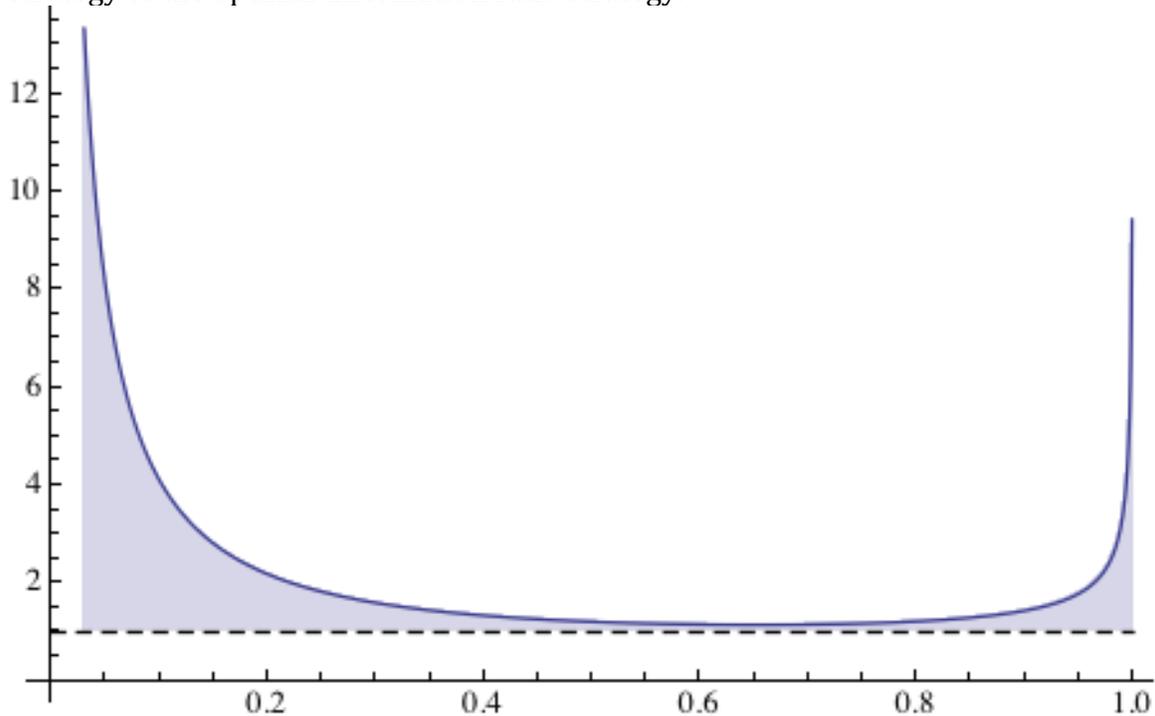

Similarly, each optimal strategy's information ratio is independent of $\sigma$. The excess of the optimal information ratio strategy's information ratio to that of the optimal expected return strategy is shown in Figure 9. The difference is small unless the correlation approaches one, in which case the difference tends to infinity.

**Figure 9: The Difference between Information Ratios of Optimal Strategies**
This figure plots the difference between the information ratio of the optimal information ratio strategy and that of the optimal expected return strategy.

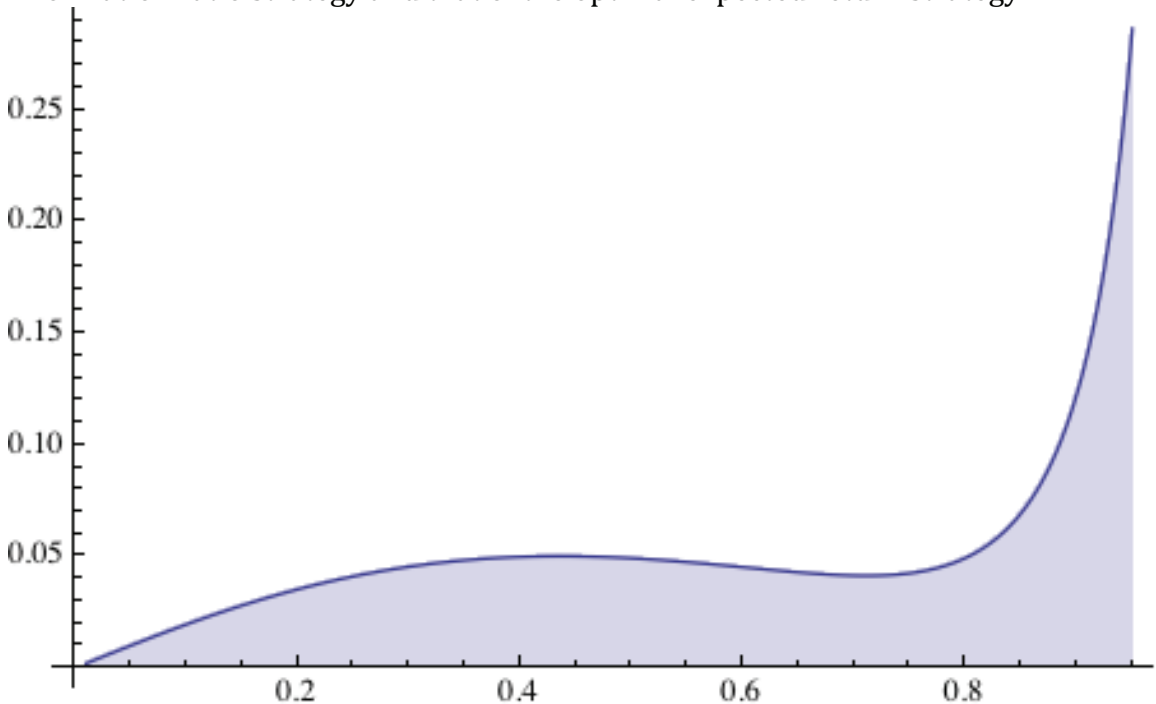

## Conclusion

How close to the best is a given trading strategy? The traditional approach in developing strategies has been to tinker and improve, never knowing how close to the optimal one is.

We answer this question instead by introducing a new general framework that explicitly constructs the optimal trading strategy for a given security and historical indicator and evaluates its expected return. We also compare the results with those of the trading strategy with the highest information ratio and find that in most real-world situations the differences are small.

Thus, we can know both how far away a proposed strategy is from the optimal, and what the optimal trading strategy itself is. This approach helps compare different strategies using different parameters and implementations across different indicators and markets.

Future research could include exploring the results in the presence of stop-loss trading strategies, the trade-off between optimality and stability particularly in cases of negligible correlations, and empirical applications to compare with standard strategies.